\documentclass[fleqn,usenatbib]{mnras}

\usepackage{newtxtext,newtxmath}

\usepackage[T1]{fontenc}

\DeclareRobustCommand{\VAN}[3]{#2}
\let\VANthebibliography\thebibliography
\def\thebibliography{\DeclareRobustCommand{\VAN}[3]{##3}\VANthebibliography}

\usepackage{graphicx}	
\usepackage{CJK}        
\usepackage[utf8]{inputenc}     
\usepackage{xspace}     
\usepackage{siunitx}    
\usepackage{bpchem}     
\usepackage{varioref}   
\usepackage{pdflscape}  
\usepackage{rotating}   

\labelformat{chapter}{Chapter~#1}
\labelformat{section}{Section~#1}
\labelformat{appendix}{Appendix~#1}
\labelformat{subsection}{Section~#1}
\labelformat{subsubsection}{Section~#1}
\labelformat{figure}{Figure~#1}
\labelformat{subfigure}{Figure~\thefigure #1}
\labelformat{table}{Table~#1}
\labelformat{equation}{Equation~(#1)}

\newcommand{\varalpha}{\TextOrMath{\(\Delta\alpha/\alpha\)}{\Delta\alpha/\alpha}\xspace}
\newcommand{\sigsys}{\TextOrMath{$\sigma_{**}$}{\sigma_{**}}\xspace}
\newcommand{\Teff}{\TextOrMath{T$_\textrm{eff}$}{\textrm{T}_\textrm{eff}}\xspace}
\defcitealias{Berke2022b}{B22b}
\DeclareSIUnit\bar{bar}

\hypersetup{breaklinks=true}

\title[Probing $\alpha$ with solar twins: methods \& results]{Probing Galactic variations in the fine-structure constant using solar twin stars: methodology and results}

\author[D. Berke et al.]{
Daniel A. Berke,$^1$\thanks{\textit{Gemini Observatory North, 670 N. A$\!$`oh\=ok\=u Place, Hilo, HI 96720, USA} Email: daniel.berke@noirlab.edu}
Michael T. Murphy,$^1$
Chris Flynn,$^1$
Fan Liu (刘凡)$^1$
\\
$^{1}$Centre for Astrophysics and Supercomputing, Swinburne University of Technology, Hawthorn, VIC 3122, Australia
}

\date{Accepted 2022 August 23. Received 2022 August 3; in original form 2022 April 1}

\pubyear{2022}

\begin{document}
\begin{CJK*}{UTF8}{gbsn}
\label{firstpage}
\pagerange{\pageref{firstpage}--\pageref{lastpage}}
\maketitle

\begin{abstract}
The rich absorption spectra of Sun-like stars are enticing probes for variations in the fine-structure constant, $\alpha$, which gauges the strength of electromagnetism.
While individual line wavelengths are sensitive to $\alpha$, they are also sensitive to physical processes in the stellar atmospheres, which has precluded their use so far.
Here we demonstrate a new, differential approach using solar twins: velocity separations between close pairs of transitions are compared across stars with very similar physical properties, strongly suppressing astrophysical and instrumental systematic errors.
We utilise 423 archival exposures of 18 solar twins from the High-Accuracy Radial velocity Planetary Searcher (HARPS), in which calibration errors can be reduced to $\lesssim$\SI{3}{\meter\per\second}.
For stars with $\approx$10 high signal-to-noise ratio spectra ($\ge$200 per pixel), velocity separations between pairs are measured with $\approx$\SI{10}{\meter\per\second} statistical precision.
A companion paper assesses a range of systematic error sources using 130 stars, with a greater range of stellar parameters, providing accurate corrections for astrophysical effects and a residual, intrinsic star-to-star scatter of 0--\SI{13}{\meter\per\second}.
Within these uncertainties, we find no evidence for velocity separation differences in 17 transition pairs between solar twins.
In a second companion paper, this is found to limit local ($\lesssim$\SI{50}{pc}) variations in $\alpha$ to $\approx$50 parts per billion, $\sim$2 orders of magnitude less than other Galactic constraints.
\end{abstract}




\section{Introduction}
The current theoretical framework of physics, underpinned by the Standard Model and the general theory of relativity, can account for a remarkable range of physical phenomena on scales from the sub-atomic to that of the visible universe.
However, this theoretical framework relies on a set of dimensionless numbers known as the ``fundamental constants,'' whose values are still fundamentally mysterious.
Their values cannot be computed or explained within the theories in which they appear, but can only be measured in nature.
\citet{Milne1935, Milne1937} and \citet{Dirac1937, Dirac1938} first raised the question of whether fundamental constants might vary in the low-energy universe, yet nearly a century later the question remains unanswered.
In the absence of an accepted theory which explains these constants' values, searches for variation in them remain a promising avenue towards the discovery of new physics.

In this paper, we detail a new approach -- the solar twins method -- for searching for variation in the fine-structure constant, $\alpha$.
We present an overview of our methodology using archival data from the High Accuracy Radial-velocity Planet Searcher (HARPS) spectrograph and the initial results of our analysis.
In a companion paper, \citet[][hereafter \citetalias{Berke2022b}]{Berke2022b}, we give a thorough overview of systematic errors in the method and our approaches to avoiding or suppressing them, while another companion paper, \citet{Murphy2022b}, provides the first constraints on variation in $\alpha$ derived from this method.

The fine-structure constant, \(\alpha\equiv e^2/\hbar c,\) was introduced by Sommerfeld in 1916 \citep{Sommerfeld1916} and is the dimensionless coupling constant for the electromagnetic interaction.
\citet{Jordan1937} first proposed that $\alpha$ might change with cosmic time, and multiple hypothetical extensions of the Standard Model have been proposed since to explain how $\alpha$ could vary, as a function of e.g., cosmic time \citep{Brans1961, Forgacs1979, Marciano1984} or dark matter background density \citep{Olive2002, Davoudiasl2019}.
Experimental searches for variation in the fine-structure constant have been underway since the 1950s \citep{Savedoff1956, Wilkinson1958} using a variety of methods and probes \citep[see, e.g.][or see \citealt{Uzan2011} or \citealt{Martins2017} for recent reviews]{Stein1974, Shlyakhter1976, Avelino2001, Murphy2001a, Nollett2002, Olive2004, Gould2006, Rosenband2008, Landau2010, Probst2014, Kotus2017, Murphy2017, Bainbridge2017a, Hees2020, Bora2020}.

The most precise methods for searching for variation in $\alpha$ involve spectroscopy, making use of $\alpha$'s role as the coupling constant for electromagnetic interactions.
Experiments with atomic clocks over the course of a few years have placed strong constraints on variation with time in the present day on Earth, at the level of
\SI[separate-uncertainty=true]{1.0+-1.1e-18}{{yr}\tothe{-1}}
\citep{Lange2021}.
However, variation on other time and distance scales in the universe is much less constrained.
The most precise astronomical constraints available, derived from 29 carefully calibrated measurements of quasar absorption systems, have a weighted mean relative variation at the level of $\langle\varalpha\rangle_{\text{W}}=-0.5\pm0.5_\mathrm{stat}\pm0.4_\mathrm{sys}\times10^{-6}$ \citep[\,parts per million, or ppm;][]{Murphy2022}.
The Galactic scale, in contrast, remains relatively unexplored, despite being a promising location to test theories of $\alpha$ varying as a function of dark matter background density \citep{Olive2002, Davoudiasl2019}.
The solar twins method is thus well-positioned to take advantage of this opportunity to probe \varalpha on the Galactic scale.

Spectroscopic searches are sensitive to variation in $\alpha$ because each electron energy level has its own dependence on $\alpha$, due mainly to the coupling between orbital and spin angular momentum, plus other relativistic effects \citep{Dzuba1999a}.
A change in $\alpha$ would induce a shift in all energy levels of an atom simultaneously, with the magnitude and sign of the change unique to each level.
The energy differences between levels would change accordingly, rendering variation in $\alpha$ detectable via the change in wavelength of photons emitted or absorbed in transitions between levels.

By comparing the wavelengths of atomic transitions between different sources, constraints on any difference in $\alpha$ between them can be measured.
For astronomical searches, variation in the fine-structure constant is defined as
\begin{equation}
\frac{\Delta\alpha}{\alpha}\equiv\frac{\alpha_\text{obs}-\alpha_0}{\alpha_0},
\end{equation}
where \(\alpha_0\) is the current laboratory value of \(\alpha\) and \(\alpha_\text{obs}\) is the value in the object used as a probe.
Quasar absorption systems have been used as probes in searches for variation in $\alpha$ since the work of \citet{Bahcall1967a}, by comparing transition wavelengths in them with their laboratory-measured values.
These systems provide the current best direct astronomical constraints on \varalpha, but each system typically contains at most $\lesssim$\,10 strong transitions, and current \SI{10}{\meter}-class telescopes have mostly exhausted the precision available from them without prohibitively long observing times \citep{Kotus2017}.

Stars, with potentially thousands of absorption features in their spectra, have long been viewed as highly promising probes of \varalpha.
Stars brighter than all but the brightest quasars are numerous and can be used to probe \varalpha on the Galactic scale, which remains relatively unexplored and would allow testing of theories relating variation in $\alpha$ to dark matter density.
However, use of stars has been limited by the physics in stellar atmospheres, which is a source of significant systematic errors that strongly limit the accuracy and precision of constraints measured using current methods.
Absorption features in stars are known to be asymmetric due to convection in their photospheres \citep[e.g.,][]{Dravins1982, Dravins2008, Gray2010a, Gray2010b, Sheminova2020}.
Changes in radial velocity caused by the difference in surface area and temperature covered by upwelling and downfalling material on the star's surface can shift the wavelength measured for transitions by hundreds of \si{\meter\per\second} from their laboratory values.
In addition, the orientation of the star's rotation axis, changes in temperature, metallicity, and surface gravity, or weak blends of differing strength between stars (from e.g., elemental abundance or isotope ratio differences) will also affect the measured wavelengths.

Previous attempts to use stars as probes of \varalpha with conventional methods have been unable to overcome these systematic issues.
Recent work by \citet{Hu2020} using observations of a single white dwarf reported two values of $\varalpha=63.6\pm3.3_\text{stat}+19.4_\text{sys}$~ppm and $\varalpha=42.1\pm4.7_\text{stat}+23.5_\text{sys}$~ppm.
The two values were calculated by comparing to two independent sets of laboratory wavelength measurements, illustrating the critical importance of laboratory measurements to current methods.
Indeed, \citet{Hu2020} state that the overall uncertainty in their result is dominated by laboratory wavelength errors.
Other recent work by \citet{Hees2020} used 5 red-giant stars as probes, with a reported constraint of $\varalpha=1.0\pm5.8$~ppm.
However this value is dominated by a single star, with the other 4 stars providing much less precise measurements at the level of 100 ppm.
In addition, important systematic errors such as the aforementioned line shifts and asymmetries were not considered in the analysis; these are likely similar in magnitude to the statistical uncertainty, i.e. 5--10~ppm.

Stars remain tantalizing probes, however, and in this work we describe a new approach to constrain \varalpha using them. 
Our solar twins method uses a differential approach designed to avoid and suppress the systematic effects encountered with current methods.
The method compares the velocity separation of the same pairs of transition between stars with intrinsically similar spectra.
This avoids any comparison with laboratory wavelengths and avoids most of the effects of line shifts and asymmetries.
Using stars all very similar to each other allows us to model and account for any remaining small differences in transition pair separations due to their slightly different atmospheric physics.
This paper, in concert with \citetalias{Berke2022b}, details the solar twins method and provides an analysis of systematic errors associated with it.
\citet{Murphy2022b} contains the first constraints on \varalpha obtained using this method, which are at the level of $\sim10^{-8}$, two orders of magnitude more precise than the current best published constraints.

While in principle any kind of star could be used as probes, Sun-like stars -- G-type dwarfs on the main sequence -- present some advantages.
They contain thousands of intrinsically narrow absorption features in their spectra, but not so many that they blend together and blanket each other as in lower-mass stars.
We can also utilize our knowledge of the Sun to gain better insight into the likely atmospheric conditions in which the lines are formed. 
Therefore, in this paper we focus even more narrowly on a set of the most Sun-like stars, known as \emph{solar twins}.

The term `solar twin,' though widely used, has no single quantitative definition in the literature, so we adopt the following definition:
\begin{equation} \label{equation:star_definitions}
    \text{Solar twin} = \left\{
    \begin{array}{r@{\,}l}
         \mathcal{T}^{\mathrm{N}}_{\mathrm{eff}\odot}&\pm\,\SI{100}{\kelvin}, \\
         \mathrm{[Fe/H]_\odot}&\pm\,\SI{0.1}{dex},\\
         \log{g_\odot}&\pm\,\SI{0.2}{dex},\\
    \end{array} \right. \\
\end{equation}
where $\mathcal{T}^{\mathrm{N}}_{\mathrm{eff}\odot}=\SI{5772}{\kelvin}$ is the nominal solar effective temperature as defined by International Astronomical Union (IAU) 2015 Resolution B3 \citep{Prsa2016}, [Fe/H] is 0 by definition, and $\log{g_\odot}=4.44$ is the solar surface gravity\footnote{Calculated as $g_{\odot}=\mathcal{(GM)}^\mathrm{N}_\odot/\mathcal{R}^2_\odot$, where $\mathcal{(GM)}^\mathrm{N}_\odot$ is the nominal solar mass parameter and $\mathcal{R}^2_\odot$ the nominal solar radius similarly defined by the IAU \citep{Prsa2016}. Note that the uncertainties on these values for the Sun are all zero, as they are either defined directly by the IAU or calculated from other defined quantities.} (for $g_\odot$ in units of \si{\centi\meter\per\second\squared}; we use the same units for all measurements of $g$ in this work).
Solar twins are of interest across multiple areas of astronomy, as they allow us to calibrate theories of star formation, structure, composition, and activity against the Sun.
Searches for solar twins have been performed since at least the work of \citet{Hardorp1978}.
In this paper we focus on the Sun and 17 solar twins in the solar vicinity (\SI{<50}{pc}), though in \citetalias{Berke2022b} we make use of an additional 112 stars (still similar to the Sun) to model pair separation changes as a function of stellar atmospheric parameters.
Future work will be able to take advantage of \textit{Gaia} to discover solar twins much further away, up to 2--4~\si{kpc}, allowing testing for variation in $\alpha$ on the Galactic scale as a function of dark matter density.
Initial progress in this area is reported in \citet{Lehmann2022}, Lehmann et al. (in prep.), and Liu et al. (in prep.).

This paper is organized as follows: in \ref{firstpaper:section:methods} we explain the solar twins method in detail.
\ref{firstpaper:section:data} describes details of the HARPS instrument and processing of data products from it.
In \ref{firstpaper:section:analysis} we give an overview of the analysis process and a brief summary of systematic errors, with a full systematic error analysis derived from a larger sample of stars presented in \citetalias{Berke2022b}.
We present the results in \ref{firstpaper:section:results}, and conclude in \ref{firstpaper:section:conclusions}.

\section{The solar twins method} \label{firstpaper:section:methods}
Here we introduce the `solar twins method' for searching for variation in $\alpha$.
This method can be summarised as comparing the separation of selected pairs of transitions with each other across very similar Sun-like stars (solar twins).
Atomic physics suggests that each transition pair's separation would respond to a variation in $\alpha$ to a different degree, both in magnitude and sign (\citealt{Dzuba1999a, Dzuba2022}).
A given change in $\alpha$ would therefore imprint a unique pattern of pair separation changes onto a spectrum.
Evidence of such a pattern of changes between different stars could then be used to infer the change in $\alpha$ necessary to produce it.
By comparing the separation between pairs of transitions across very similar stars, we avoid many of the systematic errors that have affected previously-used methods.
The use of solar twins also allows us to leverage our knowledge of physical properties in the atmosphere of the Sun (the star we can study in the closest detail) which might impart systematic effects on observations.

To apply the solar twins method, we require the following components.
First, a sample of stars similar enough to each other to allow phenomenological modelling of pair separation differences between stars due to slightly differing stellar atmospheric parameters.
The existence of these differences and the process of correcting for them are discussed in \ref{section:pair_separation_offsets}, while the stellar sample selection is discussed in \ref{firstpaper:section:stellar_sample_section}.
The sample of transitions is discussed in \ref{firstpaper:section:selecting_transitions}, and the sample of exposures are described in \ref{firstpaper:section:data_sample}. 
Second, we need conversion factors between pair separation changes and the corresponding change in $\alpha$.
We discuss this aspect in \ref{firstpaper:section:many_multiplet_method}.
Finally, a rigorous analysis of systematic errors is required in order to be confident that changes are due to $\alpha$ rather than other causes (such as stellar atmospheric differences).
A full explanation of systematic errors identified in the process of this analysis is presented in \citetalias{Berke2022b}, but we include a summary of 12 effects we considered in \ref{firstpaper:section:summary_of_systematic_errors}.

\subsection{Relating pair separation changes to variation in the fine-structure constant} \label{firstpaper:section:many_multiplet_method}
The energy of each electron orbital state depends on $\alpha$ in its own characteristic way.
A change in $\alpha$ would cause the energy of a transition between two states to change, with $\Delta E/E \propto \Delta \alpha/\alpha$ (for small $\Delta\alpha/\alpha$).
The proportionality constant for each transition is known as a $q$-coefficient \citep{Dzuba1999a, Webb1999}.
The change in energy translates to a velocity shift in the transition, $\Delta v/c \propto \Delta E/E$, with the magnitude and sign given by the $q$-coefficient.
For $\Delta\alpha/\alpha\ll1$ (known from previous searches for variation), the relationship can be approximated as a linear function and quantified by \citep[e.g.,][]{Murphy2017}:
\begin{equation}
    \frac{\Delta v}{c}\approx-2\frac{\Delta\alpha}{\alpha}\frac{q}{\omega},
\end{equation}
where $\omega$ is the wavenumber of the transition in \si{\per\centi\meter} ($q$ has the same units).
Importantly, by construction, an incorrect $q$-coefficient will not lead to a spurious detection of a non-zero change in $\alpha$ where none exists \citep{Murphy2001a, Murphy2003}.

For an order of magnitude estimate of the velocity shifts, consider a transition at \SI{500}{\nano\meter} with a representative $q$-coefficient of +\SI{1000}{\per\centi\meter}.
This is typical of $q$-coefficients for the transitions utilized here, which are listed in \ref{table:transitions} for reference, from \citet{Dzuba2022}.
For a change in $\alpha$ of 1 part in $10^{6}$ (the precision level of current astronomical tests), the change in velocity space is approximately \SI{-30}{\meter\per\second}.
Improving on current astronomical constraints therefore requires reaching at least this level of precision and accuracy.

\subsection{Selection of transitions} \label{firstpaper:section:selecting_transitions}
Prior to this work, $q$-coefficients were unavailable for optical (rest frame) transitions observed in main-sequence stars.
The difficulty of calculating $q$-coefficients varies with transition and in general is non-trivial (V. Dzuba, priv. com.), so, as we did not know beforehand how many would prove viable, our first step was to assemble as large a sample as possible of observationally-usable transitions from which to draw.
The full selection process is detailed in \citetalias{Berke2022b}, but in summary, we selected transitions from a list of 8843 transitions (A. Lobel,  priv. comm.) from the Belgian Repository of fundamental Atomic data and Stellar Spectra \citep[BRASS,][]{Laverick2017}, which matched transitions in simulated spectra to solar features.
To acquire the electron orbital configurations of these transitions, we cross-matched them in the Atomic Spectra Database \citep{Kramida1999} of the National Institute of Standards and Technology (NIST).
To reduce errors from saturated or weak lines, the selection of transitions was limited to those with normalized depths relative to the continuum of between 0.15 and 0.90, as measured in a HARPS solar spectrum (SNR\,=\,317) reflected off the asteroid Vesta.
After selecting the least visually-blended transitions, we were ultimately left with a sample of 164 potentially usable transitions.
Of these, 22 transitions were able to have $q$-coefficients calculated for them \citep{Dzuba2022}, and these 22 transitions constitute the ``transition sample'' for this paper.
Information on these transitions including wavelengths, atomic species, orbital configuration of upper and lower states, and their $q$-coefficients is listed in \ref{table:transitions}.

Of the 12 sources of systematic error we considered for this analysis (listed in \ref{firstpaper:section:summary_of_systematic_errors}), two important sources that are difficult to account for are strong blending of absorption features with either telluric or other stellar absorption features.
Such blending could add hundreds of \si{\meter\per\second} to our measurements.
As selecting transitions to preemptively avoid blending strongly defined the final transition sample, we discuss these sources of systematic error in more detail individually in the next two subsections.

\subsubsection{Blending with telluric features} \label{firstpaper:section:telluric_features}
The radial velocity between an Earth-bound observer and any given star changes both daily due to the Earth's rotation and yearly due to the Earth's annual motion around the barycentre of the solar system.
This causes the star's spectrum to shift in the observatory's rest frame as a function of time.
As part of its automated data reduction process, HARPS includes a barycentric correction with every spectrum to account for the observatory's motion relative to the target star.
Telluric absorption features, however, are independent of the Earth's annual motion and remain at the same wavelengths throughout the year, appearing at different locations relative to the stellar spectrum over time.
A telluric absorption feature blending with a nearby stellar absorption feature introduces significant systematic errors in the measurement of the stellar feature's wavelength for the measurement process we used (described in \ref{firstpaper:section:automated_fitting}).
The problem is compounded by the degree of blending varying with the phase of the year rather than being fixed.

The change in Earth's barycentric radial velocity has a maximum peak-to-peak amplitude of \SI{\sim60}{\kilo\meter\per\second}, so any spectral absorption feature within $\sim$\SI{30}{\kilo\meter\per\second} of a telluric feature can potentially be blended with it depending on the date of observation\footnote{The maximum change in the Earth's radial velocity over the course of a year is $\sim$\SI{11}{\centi\meter\per\second\per\minute}, so given the maximum exposure time of $\sim$\SI{1}{\kilo\second} for the exposures in our sample, we derive an upper limit of $\sim$\SI{2}{\meter\per\second} change in a single exposure. While this smears all spectral features by this amount, this is below the systematic noise floor of \SI{\sim6}{\meter\per\second} discussed in \ref{section:star_to_star_scatter} and we may safely neglect it.}.
In addition, the barycentric radial velocities of stars in the sample must be considered or a feature may be safely unblended in one star yet blended in another.
We investigated this issue in detail in \citetalias{Berke2022b} and ultimately selected only transitions greater than \SI{100}{\kilo\meter\per\second} away from telluric features deeper than 0.1\% of the continuum.
These features were found using a synthetic atmospheric transmission spectrum from the Transmissions Atmosph\'eriques Personnalis\'ees Pour l'AStronomie online service\footnote{\url{http://ether.ipsl.jussieu.fr/tapas/}}, calculated at the location of La Silla Observatory \citep{Bertaux2014}.
This restriction on proximity to telluric features allows a range of barycentric radial velocities of \SI{\pm70}{\kilo\meter\per\second} for stars while still avoiding blending from the annual \SI{\pm30}{\kilo\meter\per\second} change.
As calculated in \citetalias{Berke2022b}, any remaining error in the centroid of a single line from blending should be less than \SI{\sim30}{\meter\per\second}.
This selection left us with 783 transitions from the initial 8843.

\subsubsection{Blending with other stellar features} \label{firstpaper:section:blending_stellar_features}
In addition to telluric features, absorption features in solar twins are often blended with other nearby stellar features.
We visually inspected each transition in the sample above in the high-SNR solar spectrum (Vesta) and rated each transition on a scale of 0 (``no blending apparent'') to 5 (``extremely blended'') based on profile shape and asymmetry.
From this assessment we elected to use only transitions in categories 0 to 2, which left us with the selection of 164 transitions in \citetalias{Berke2022b}.
Examples of feature profiles in each category and an analysis of the scatter and mean error in measured wavelengths for transitions in each category can be found as figures 3 and 9 in \citetalias{Berke2022b}.

\subsection{Selection of transition pairs} \label{firstpaper:section:selecting_pairs}
Pairs of transitions were carefully selected to avoid two possible causes of systematic error.
First, we required that both absorption features in a pair differ in normalized depth (relative to the continuum) by no more than 0.2.
Absorption features form over a range of (physical) depths in stellar photospheres, which impart shifts and asymmetries to their profiles due to the different physical conditions along the way \citep{Dravins1982,Dravins2008}.
From a purely phenomenological standpoint, features with similar normalized depths tend to exhibit similar asymmetries and, as stellar parameters vary across stars, lines with similar depth with tend to change in similar ways, so their separation will be a simpler, easier-to-model function of those parameters.
We tested for evidence of systematic error as a function of the difference in normalized depth between features in a pair, and found none with the 0.2 limit we imposed (as seen in figure 10 of \citetalias{Berke2022b}).

Second, the HARPS spectrograph (see \ref{firstpaper:section:harps}) is known to have intra-order distortions in its wavelength calibration scale of up to \SI{\pm50}{\meter\per\second} across an echelle order \citep{Molaro2013}.
While we have attempted to correct for these distortions (see \ref{firstpaper:section:analysis} for details), we also limited the maximum velocity separation between transitions in a pair to \SI{800}{\kilo\meter\per\second} in order to help limit possible systematic errors from this source.
To avoid pairs of transitions blended at HARPS's resolutions, we also imposed a minimum separation limit of \SI{9.1}{\kilo\meter\per\second}, 3.5 times HARPS's resolution element \citep{Mayor2003}.
While we were able to find 229 pairs fitting these criteria, in this paper we focus on the 17 pairs where both transitions had $q$-coefficients available.
These 17 pairs constitute the ``pair sample'' for this paper.
\ref{table:transitions} contains details on which pair each transition in the transition sample is part of.

\subsubsection{`Instances' of transitions and transition pairs}
Because HARPS has echelle orders with overlapping spectral range (described in \ref{firstpaper:section:harps}), some transitions can be measured twice in adjacent diffraction orders.
If both transitions in a pair fall within an overlapping region, pairs can also be measured twice.
This provides a prime opportunity to test for systematics across HARPS's detector.
When a pair or transition is measured twice, we refer to the two measurements as ``instances'' and treat them individually for the purpose of systematic error checking, using the echelle order number to differentiate them.
Three pairs in the pair sample have two instances in this manner, plus an additional 52 pairs in the broader sample of \citetalias{Berke2022b}.

\subsection{Stellar sample selection} \label{firstpaper:section:stellar_sample_section}

\begin{table}
\centering 
    \begin{tabular}{ |l|l|r|c| r @{/} l | }
    \hline
    Star & \Teff & [Fe/H] & $\log g$ & \multicolumn{2}{c|}{obs.}\\
    \hline
    Sun & 5772 & 0 & 4.44 & 22 &0\\
    HD 1835 & 5747 & 0.08 & 4.45 & 1&0\\
    HD 19467 & 5753 & $-0.07$ & 4.30 & 7&0\\
    HD 20782 & 5773 & $-0.09$ & 4.39 & 8&6\\
    HD 30495 & 5857 & $-0.02$ & 4.50 & 2&0\\
    HD 45184 & 5863 & 0.04 & 4.42 & 104&7\\
    HD 45289 & 5710 & 0.03 & 4.25 & 12&0\\
    HD 76151 & 5787 & 0.03 & 4.43 & 7&0\\
    HD 78429 & 5740 & 0.05 & 4.27 & 19&33\\
    HD 78660 & 5788 & $-0.03$ & 4.39 & 1&0\\
    HD 138573 & 5745 & $-0.04$ & 4.41 & 0&1\\
    HD 140538 & 5693 & 0.05 & 4.46 & 11&0\\
    HD 146233 & 5826 & 0.06 & 4.42 & 91&54\\
    HD 157347 & 5730 & 0.03 & 4.42 & 15&4\\
    HD 171665 & 5725 & $-0.10$ & 4.46 & 3&0\\
    HD 183658 & 5824 & 0.06 & 4.46 & 4&0\\
    HD 220507 & 5701 & 0.02 & 4.26 & 2&8\\
    HD 222582 & 5802 & 0.00 & 4.25 & 1&0\\
    \hline
    \end{tabular}
    \caption{
    Stellar parameters for the stars considered in this paper, with columns showing each star's identifier, effective temperature, metallicity, surface gravity, and the number of observations before/after the HARPS fibre change in 2015 (instrumental differences necessitating this split, see \ref{firstpaper:section:harps_fiber_change}). 
    All values are from \citet{Casagrande2011} except for the solar values, which are from \citet{Prsa2016}.
    }
    \label{table:stars}
\end{table}

For selecting stars we used the Geneva--Copenhagen Survey (GCS), a sample of 14139 nearby F and G dwarfs \citep{Nordstrom2004}.
We used stellar parameters from \citet{Casagrande2011}, who derived updated photometric values of \Teff, [Fe/H], and $\log{g}$ for the stars in the GCS.
In this paper we focus on 18 solar twins from  a larger collection of 130 stars described in \citetalias{Berke2022b}.
Stars in this larger collection were selected over a wider range of the stellar atmospheric parameters \Teff, [Fe/H], and $\log{g}$ than used for solar twins in order to model pair separation changes as a function of those parameters.
While it is tempting to add further solar twins to the sample from more recent, more targetted studies \citep[e.g.,][]{Bedell2018, Galarza2021}, by limiting our analysis in this paper to just the solar twins in that larger collection, we obtain the most homogeneous sample from which to derive the first reported results from the new solar twins method (while still making use of the modelling described in \citetalias{Berke2022b}).

\ref{table:stars} shows the 18 solar twins that make up the ``stellar sample'' for this paper, along with their effective temperature, metallicity, surface gravity, and number of observations.
Note that this sample includes the Sun, which we treat the same as the other stars; per \ref{equation:star_definitions}, the Sun \emph{itself} is a solar twin, so we make use of this technicality to avoid specifying it separately all the time.
While spectroscopically-derived parameters are available in \citet{Spina2020} for some stars in the larger \citetalias{Berke2022b} collection, we compared the values from \citet{Casagrande2011} for 16 stars common to both and found no significant systematic differences.
As these spectroscopically-derived values are not available for all stars in the \citetalias{Berke2022b} collection, we chose to use the photometric values in order to have a homogeneously-derived set of stellar parameters.

\section{Spectrograph and data} \label{firstpaper:section:data}
A more detailed description of the data analysis process used in this analysis can be found in \citetalias{Berke2022b}, but we provide a summary here.

\subsection{The HARPS spectrograph} \label{firstpaper:section:harps}
For this work we used archival spectra from the High-Accuracy Radial velocity Planetary Searcher (HARPS), located on the European Southern Observatory's (ESO) \SI{3.6}{\meter} telescope at La Silla Observatory, Chile \citep{Pepe2002, Mayor2003}.
HARPS is a high resolution ($R=115000$), cross-dispersed echelle spectrograph covering most of the visible range from 380 to \SI{690}{\nano\meter} \citep{Mayor2003}.
Its primary science goal is to search for exoplanets using the radial velocity technique, so it is designed for high-precision spectroscopy with an accurate and, most importantly, stable wavelength scale.
It has been used to observe many Sun-like stars as part of its exoplanet-hunting campaigns, with thousands of archival spectra available from 2003 onward.
While the requirements for planet searching are not quite the same as our requirements for precise, accurate measurements of individual absorption features, HARPS's stability has allowed very detailed characterization and calibration of its wavelength scale \citep{Wilken2010a,Bauer2015,Coffinet2019,Milakovic2020}.

To minimize instrumental drift, HARPS is encased in a vacuum chamber ($\mathrm{pressure}<\SI{0.01}{\milli\bar}$) which is kept temperature stabilised with an expected long-term stability of \SI{0.01}{\degreeCelsius} \citep{Mayor2003}.
To combat nightly drifts HARPS uses a simultaneous calibration scheme, using two fibres to observe both the target and a reference spectrum simultaneously on separate portions of the same detector \citep{Mayor2003}.
HARPS's echelle grating gives it a high resolution while still allowing a compact design by dispersing incoming light into a number of echelle orders which are then cross-dispersed across the detector.
Adjacent orders contain some overlap in their wavelength ranges, which is largest at the blue end and decreases to almost none at the red end.
In normal operation the automated HARPS pipeline \citep{Rupprecht2004} merges the overlapping ends of each order to create a single, 1D output spectrum.
For reasons explained in \ref{firstpaper:section:uncertainties_array}, however, we use the data products from an intermediate stage of the pipeline, after extraction of all orders but prior to merging.
This has the additional effect of allowing consistency checks by comparing measurements  of the same features taken on opposite sides of the detector in adjacent orders.

\subsubsection{Change of optical fibres in HARPS} \label{firstpaper:section:harps_fiber_change}
Between 19 May--3 June 2015 HARPS was upgraded with new, octagonal optical fibres to replace its original circular ones \citep{LoCurto2015}.
This upgrade caused a change in the instrumental profile of the spectrograph, which propagated into a shift in the measured wavelengths of absorption features before and after the change \citep{LoCurto2015, Dumusque2018}.
For this reason, we treat observations from the two eras before and after the fibre change as if from separate instruments, and only compare measurements taken in the same era with each other.
We refer to these periods of time as the `pre-' and `post-' fibre-change eras.

\subsection{Data used} \label{firstpaper:section:data_sample}
For this project we used archival HARPS spectra from the ESO Science Archive Facility\footnote{\url{http://archive.eso.org/wdb/wdb/adp/phase3\_spectral/form}} spanning a range of observation dates from 2004 to 2017.
To minimise the effects of photon noise and enable a thorough exploration of systematic effects in the solar twins method, we chose to use only very high SNR spectra, $\mathrm{SNR}\geq200$.
Visual inspection of spectra with $\mathrm{SNR}\gtrsim450$ showed artefacts which we interpreted as signs of CCD saturation.
To avoid these artefacts we used only spectra with $200\leq\mathrm{SNR}\leq400$, except for the Sun, where we decided to decrease the lower limit to 150 due to there being just 6 solar spectra with SNR $>200$ in the archive.
With this new limit a total of 22 observations of the Sun were obtained from reflection spectra of the asteroid Vesta.
As described in \ref{firstpaper:section:harps_calibration} we also only used observations on nights for which improved wavelength calibration information was available.
A total of 423 spectra were available for the stars analysed in this paper, with the number of spectra for each star listed in \ref{table:stars}.

\subsection{Data reduction} \label{firstpaper:section:data_reduction}
HARPS has a dedicated, automated data reduction pipeline called the Data Reduction System (DRS) \citep{Rupprecht2004}.
The DRS extracts and calibrates the echelle spectra, then stitches the orders together into a single, 1D spectrum.
This 1D spectrum is included, along with intermediate data products, in the data products available in the HARPS archive.
A complete description of the data reduction process used is given in \citetalias{Berke2022b}, but we outline the important details for understanding this paper here.

\subsubsection{Creation of uncertainty arrays} \label{firstpaper:section:uncertainties_array}
The 1D spectra from the DRS do not include an uncertainty array, which is necessary for our work focusing on individual absorption features.
Following \citet{Dumusque2018}, we created uncertainty arrays by taking the square root of the quadrature sum of the flux in each pixel plus the CCD dark and read-out noise.
However, this could not be done with the 1D spectra as the individual orders are corrected for the blaze function prior to combination, removing information about the raw photon counts in each pixel.
We instead used intermediate data products provided by the HARPS archive, in the form of extracted 2-dimensional spectra (``e2ds'') files saved after optimal-extraction of each diffraction order but before any further processing is applied.
After creating uncertainty arrays as described above, we used blaze correction functions\footnote{Obtained for each night from \url{http://archive.eso.org/cms/eso-data/eso-data-direct-retrieval.html}.} to correct the flux and uncertainty arrays together.

\subsubsection{Wavelength calibration} \label{firstpaper:section:harps_calibration}
HARPS has two fibres which can receive light independently, which allows simultaneous wavelength calibration during observations.
Calibration in HARPS is normally performed by observing a calibration source at the beginning of each night with both fibres, then observing a calibration source with the `B' fibre during observations while the target is observed with the `A' fibre \citep{Mayor2003}.
The calibration source for both nightly (absolute) and simultaneous calibration at the beginning of HARPS's operations in 2004 was a thorium-argon (ThAr) lamp.
It remains in use today for absolute calibration, but individual nights since 2012 have seen increasing use of a Fabry-Perot etalon as the reference calibrator used in the B fibre.
Though HARPS has had a laser frequency comb (LFC) attached since at least the work of \citet{Wilken2010a}, it was not used to calibrate the wavelength scale of any of the spectra analysed in this work.

Over the years of HARPS's operation much work has been done to investigate and improve its wavelength calibration.
The full details of these efforts and how we incorporated their improvements into our data processing are given in Section 3 of \citetalias{Berke2022b}, though a brief summary is relevant here.
\citet{Coffinet2019} used the LFC to perform an in-depth investigation of discontinuities known to be present in HARPS's wavelength scale due to irregular pixel column widths \citep{Wilken2010a,Molaro2013,Bauer2015}.
This systematic effect produced deviations in the calibration scale of up to $\SI{50}{\meter\per\second}$ throughout the spectral range on scales of hundreds of \si{\kilo\meter\per\second}.
As part of their work they derived new calibrations which corrected for these distortions for a large number of nights.
These improved calibrations were not yet published by the time of our analysis but were generously made available for our use (C. Lovis, priv. com.).
Due to the significant improvements in accuracy from these new calibrations compared to those produced by the DRS, we used only observations for which the new calibrations were available.

\citet{Milakovic2020} reported additional systematic errors in the calibration scale at the level of \SI{10}{\meter\per\second} even after correcting for irregular pixel column widths (see Section 3 in \citetalias{Berke2022b} for more details).
They were able to map these residual intra-order distortions across the CCD by comparing the wavelength dispersion relation for the ThAr lamp with that from the LFC.
Using this map (D. Milakovi\'c, priv. com.), we were able to correct each transition based on its location on the CCD to remove the distortions.
Any remaining systematic errors in wavelength calibration after applying the corrections mentioned above are expected to be no more than a few \si{\meter\per\second}, below the typical \SI{\sim6}{\meter\per\second} noise floor from star-to-star scatter (described in \ref{section:star_to_star_scatter}).

\section{Analysis} \label{firstpaper:section:analysis}
\subsection{Automated absorption feature fitting} \label{firstpaper:section:automated_fitting}
The known shifts and asymmetry in the absorption features in stellar spectra makes comparing their wavelengths to laboratory-measured values prone to systematic error \citep[see, e.g.,][]{Dravins1982,Dravins2008}.
Differences between stellar features and their laboratory wavelengths are often hundreds of \si{\meter\per\second}.
For instance, a sample of 311 unblended \ion{Fe}{i} lines in the solar spectrum from \citet{Dravins1981} showed a range of offsets from their laboratory wavelengths of \SI{\sim1}{\kilo\meter\per\second}.
However, in the solar twins method only relative velocity separations between pairs of transitions are important, rather than comparing them to laboratory values.
Line shifts and asymmetries caused by convective turbulence simply contribute to the measured pair separation.
Importantly, the pair separation in one star is compared directly to that in similar stars rather than to any laboratory value.
This differential approach allows the solar twins method to suppress systematic errors from line asymmetries to below the level of $\sim$\SI{10}{\meter\per\second}.

In order to measure the wavelengths of absorption features in a uniform manner without human bias, we used an automated least-squares fitting technique to fit the central seven pixels in the core of each feature with a symmetrical integrated Gaussian function.
These seven pixels correspond to \SI{5.7}{\kilo\meter\per\second} at a wavelength of \SI{500}{\nano\meter} in the centre of HARPS's spectral range.
The use of a symmetrical function to fit a known asymmetrical feature will introduce some systematic error, but as explained above this error is simply incorporated into the pair separation.
In addition, asymmetries typically manifest more in the wings of features than the cores \citep{Dravins2008}, so the narrow window used for fitting helps limit any potential error from this source.

\begin{figure*}
  \begin{center}
    \includegraphics[width=\linewidth]{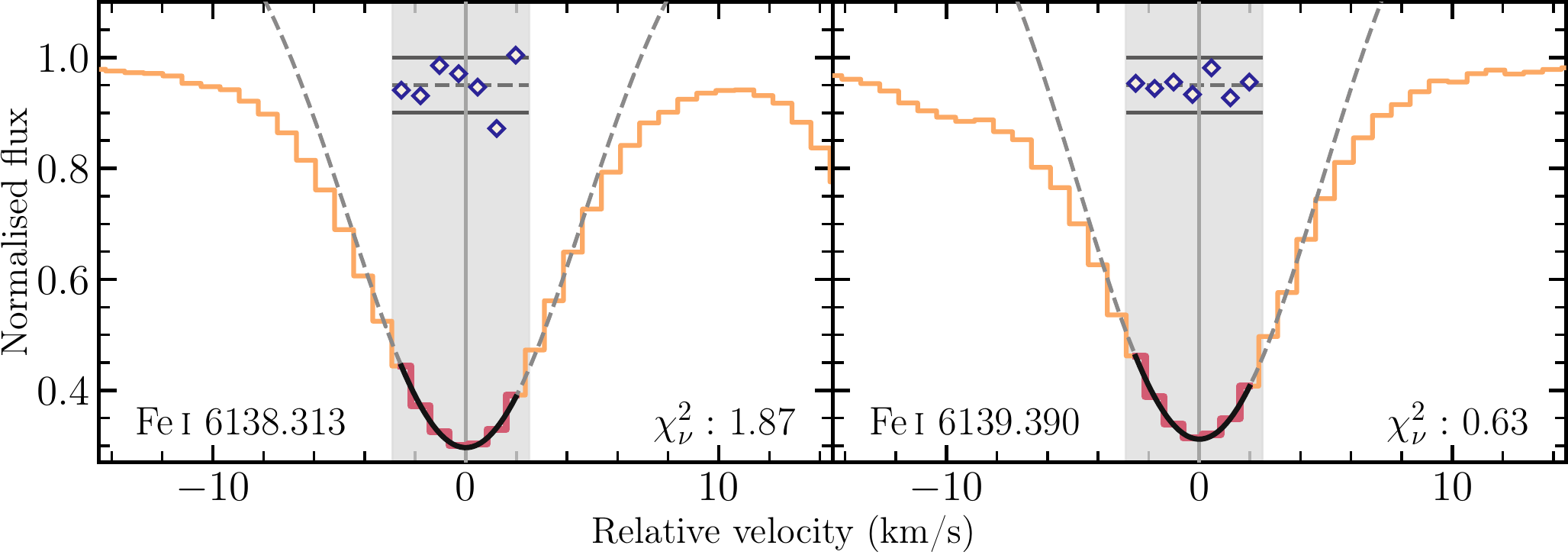}
    \caption{An example of the automated fits for the two features in one of the pairs studied, \ion{Fe}{i}\,6138.313--\ion{Fe}{i}\,6139.390.
    In each panel, the vertical line marks the fitted centroid of the feature, while the shaded vertical region illustrates the width of the seven central pixels used in creating the Gaussian fit.
    The fit itself is shown as the dashed line, with a solid black portion in the central region to emphasise which data points were fitted (the pixels in this region are also plotted a darker color than the rest of the spectrum).
    The residuals of each fit, normalized by the error array, are plotted at the top of each panel, with solid horizontal lines at $\pm1\sigma$.
    At this SNR (226 per \SI{0.82}{\kilo\meter\per\second} pixel) the errors are smaller than the thickness of the lines.
    }
    \label{figure:automated_fitting_demonstration}
  \end{center}
\end{figure*}

The fitting was performed using a custom Python package named \textsc{VarConLib}\footnote{For ``Varying Constants Library,'' code available at \url{https://github.com/DBerke/varconlib}.} developed for this work.
\textsc{VarConLib} first corrects for the radial velocity between the Earth and target using the value from the FITS header of the observation being analysed.
For each transition it then calculates the expected wavelength and takes the pixel with the lowest flux within \SI{5}{\kilo\meter\per\second} as the deepest point of the associated feature.
It then fits the following four-parameter integrated Gaussian function to the seven pixels at the core of the feature using \textsc{SciPy}'s \texttt{curve\_fit} function:
\begin{equation}
    f_n(a, b, \mu, \sigma) = \int_{\lambda_n}^{\lambda_{n+1}} \left[a\exp\left({\frac{(\lambda-\mu)^2}{2\sigma^2}}\right) + b\right]d\lambda,
\end{equation}
where $a$ and $b$ are the amplitude and baseline of the Gaussian, $\mu$ and $\sigma$ are the centroid and standard deviation, and $\lambda_n,\lambda_{n+1}$ are the wavelengths at the edges of each pixel in the dispersion direction, for $n=1,2,\dots,7$.
Note that we do not normalise the spectra before fitting, and that negative amplitudes are fit to absorption features.
We fit an integrated Gaussian to represent the total flux in each pixel because this is formally more correct than using the value of the Gaussian at the centre of each pixel as is often assumed.
The mean $(\mu)$ of the Gaussian is taken as the best-fit wavelength of the feature and the correction described in \ref{firstpaper:section:harps_calibration} is applied.
The width $(\sigma)$ of the fitted Gaussian is taken as the uncertainty for the measurement; at the SNR used in this work, the 1-$\sigma$ uncertainty for an individual feature tends to have a statistical noise floor around \SI{15}{\meter\per\second}.

\ref{figure:automated_fitting_demonstration} shows an example of the automated fits created for a pair of transitions in a solar spectrum.
As shown by the $\chi^2_\nu$ values of the residual distributions after subtracting the fitted models, the central seven pixels of these features are well-approximated by a Gaussian function even if possible asymmetries or blends are visible in their wings.
The baseline of the Gaussian is not constrained to be at the level of the continuum (which we take to be the maximum pixel value within \SI{25}{\kilo\meter\per\second}) since this is simply meant as a fast, reproducible method for measuring the centroid of the line core, rather than a representation of the entire feature.

The absorption line fitting process is applied to all selected features in all observations.
With wavelengths for all transitions established, the velocity separation, $\Delta v_\text{pair}$, between each pair of transitions selected in \ref{firstpaper:section:selecting_pairs} is calculated as the difference between the measured wavelength of each transition, $\Delta v_\mathrm{pair}=2c(\lambda_\mathrm{red}-\lambda_\mathrm{blue})/(\lambda_\mathrm{red}+\lambda_\mathrm{blue})$, where $\lambda_\mathrm{red}$ and $\lambda_\mathrm{blue}$ refer to the wavelengths of the red and blue transitions in each pair.
In the lowest-metallicity stars in our sample, some of the weakest lines selected can become undetectable even with the high signal-to-noise ratio spectra we use.
In these cases, the fitted Gaussian may have a positive amplitude from fitting a noise peak instead of a real absorption line.
Any such fits were rejected by the fitting process.

\subsection{Modelling pair separation changes as a function of stellar parameters} \label{section:pair_separation_offsets}
An important assumption of the solar twins method is that intrinsic pair separation differences in similar stars are small.
This was confirmed in \citetalias{Berke2022b}; however, absorption features' measured wavelengths were also found to vary systematically between stars as a function of their atmospheric parameters \Teff, [Fe/H], and $\log{g}$.
The process of modelling the changes for each pair so they can be compared across stars is described in detail in Section 3.5 of \citetalias{Berke2022b}, but we provide a summary below.

We used an iterative method of fitting each pair separation's weighted mean value over all exposures in each star in the larger stellar sample of \citetalias{Berke2022b}, as a multivariate quadratic function of \Teff, [Fe/H], and $\log{g}$.
The sample of \citetalias{Berke2022b} covers a wider range of the three parameters listed above, and includes the solar twin sample of this paper.
During this process the fit to the model function is iteratively optimised, with outliers redetermined and flagged at each step.

A systematic error term, \sigsys (pronounced ``sigma star-to-star''), is also added in quadrature to the statistical uncertainties and determined simultaneously during the process.
This term is a catch-all systematic error term representing additional star-to-star scatter seen in pair separations beyond what is expected from the statistical uncertainties for the individual stars.
A value of zero for a transition pair would indicate that the scatter seen in that pair's separation between stars is entirely accounted for by the statistical uncertainties.
At each step the model found is subtracted from each star's weighted mean pair separation, the $\chi^2_\nu$ value for the resulting distribution is calculated, and \sigsys is adjusted until $\chi^2_\nu$ for the distribution is unity.
The value of \sigsys for each pair then measures the remaining systematic error for that pair.
For this paper's pair sample, \sigsys ranged from 
$\sim$0--\SI{15}{\meter\per\second} for the 17 pairs considered.

These model-corrected pair separations are the primary results of the analysis, and are the difference between the measured separation in a star and the separation expected for that star based on the model.
A positive value indicates a larger difference between the component transitions in the pair than expected from the model, and vice versa.
Virtually every pair, upon visual inspection, clearly displayed a systematic variation with one or more stellar atmospheric parameters.
An example can be seen in Figure 5 of \citetalias{Berke2022b}.
The magnitude of these variations, across the parameter range studied in \citetalias{Berke2022b}, ranged from indistinguishable from statistical noise up to nearly \SI{400}{\meter\per\second}, though over the narrower parameter range of solar twins the change would generally be <\SI{100}{\meter\per\second}.
We visually inspected histograms of model-corrected separations for each pair, and they appeared normally distributed around zero in all cases.
This indicates that our choice of model (detailed in \citetalias{Berke2022b}) is adequate and effective in accounting for variations in pair separations with stellar parameters.

As changes in pair separations derive from changes in their component transitions, we first performed the same analysis to correct for systematic changes with stellar parameters and reject outliers on individual transitions.
Instead of pair separation, the quantity measured was the velocity separation between the measured wavelength and the expected wavelength (i.e., laboratory wavelength corrected for radial velocity).

\subsection{Summary of systematic errors in the solar twins method} \label{firstpaper:section:summary_of_systematic_errors}
We begin this section by noting that, when considering sources of systematic error, there are different levels where each is important.
At one level there are effects affecting individual transition measurements and pair separations made from those measurements.
For the SNR of spectra used in this work, individual measurements tend to have a statistical noise floor of \SI{\sim15}{\meter\per\second}, so systematic effects smaller than this are negligible at this level.
For placing constraints on \varalpha by comparing pair separations between stars, however, we use the weighted mean of all observations for each star.
At this inter-star level the noise floor is set by the intrinsic star-to-star scatter represented by \sigsys (as described in \ref{section:pair_separation_offsets}), with effects at the first level reduced by the square root of the number of observations.
The discovery of this intrinsic scatter is one of the important results of \citetalias{Berke2022b}, and since it varies by pair we reserve a thorough exploration of it for \ref{firstpaper:section:results}.

It is also important to note that the three largest systematic effects identified in this analysis have been accounted for and removed.
Two sources of distortion in HARPS's wavelength calibration are discussed and corrected for down to the level of, at most, a few \si{\meter\per\second} in \ref{firstpaper:section:harps_calibration}.
These distortions cause systematic effects at the individual transition level, and any remaining effects are therefore negligible.
Systematic changes in pair separations between stars, and our approach to modelling and removing them, is discussed in \ref{section:pair_separation_offsets}.
Remaining star-to-star scatter values are $\lesssim\SI{15}{\meter\per\second}$ for all pairs considered in this paper.

We briefly summarise here other potential sources of systematic error identified in our analysis, with section 4 of \citetalias{Berke2022b} containing detailed discussions of each effect.

\textbf{Transition position on the CCD:} We searched for position-dependent effects in pair separation on two main scales: opposite ends of the CCD (\SI{\sim3350}{\kilo\meter\per\second}), and over the range of radial velocities found in stars in the larger stellar sample of \citetalias{Berke2022b} (\SI{\sim200}{\kilo\meter\per\second}).
We found systematic differences in pair separation in duplicate instances of pairs found on opposite sides of the CCD, on the order of \SI{\sim20}{\meter\per\second}, but modelling each instance's dependence on stellar parameters individually effectively corrected for these differences (as described in \ref{section:pair_separation_offsets}).
We detected no evidence of systematic changes in pair separation above the level of the noise floor for individual transitions associated with the range of radial velocities found in the stellar sample.

\textbf{Absorption feature blending in stars:} Many of the absorption features may be blended with weaker transitions.
In \citetalias{Berke2022b} we visually classified how blended each absorption feature was on a scale of 0--5, and only used transitions with a ``blendedness'' of up to 2.
An analysis of pair separations by blendedness showed strongly increased root-mean-square (RMS) scatter and mean error in pair separation when a transition of blendedness 4 or 5 was involved, indicating the importance of avoiding strongly blended features.
Transitions used in this work showed a RMS of $<\SI{10}{\meter\per\second}$ over many exposures of individual stars, which should be negligible.

\textbf{Absorption feature depth:} Absorption features of different optical depths form over different ranges in stellar photospheres, and are subject to different atmospheric conditions at different heights.
This could potentially impart systematic effects as a function of the optical depth of a transition.
We found no evidence for systematic shifts as a function of mean depth of the transitions in a pair, or as a function of the depth difference between them.
Pairs comprising the deepest transitions (absorbing more than $\sim$70\% of the continuum) did show a slightly increased scatter at the level of 5--\SI{10}{\meter\per\second}, but as this is below the noise floor for individual transitions it did not warrant their exclusion.
However, future work may benefit by excluding such deep transitions or investigating their apparent increased scatter further.

\textbf{Charge transfer inefficiency (CTI):} Imperfect transfer of electrons between pixels across CCDs causes flux- and position-dependent shifts in the centroids of spectral features, and has been demonstrated in HARPS with a LFC \citep{Zhao2014, Zhao2021}.
\citet{Berdinas2016} used RV measurements of M dwarfs taken at a range of $\mathrm{SNR}\sim20$--225 to derive a formula for $\Delta\mathrm{RV}$ due to CTI in HARPS.
Even at $\mathrm{SNR}\approx20$, changes in RV were \SI{<15}{\meter\per\second}, dropping rapidly to \SI{<5}{\meter\per\second} for $\mathrm{SNR}\geq40$, suggesting no systematic shift in feature centroids above this level at our chosen SNR.
This is also below the statistical noise floor for individual transitions, and should thus be negligible as a source of systematic error.

\textbf{Lunar illumination contamination:} The solar spectrum reflecting off the Moon and scattered by the atmosphere can cause systematic errors in centroid measurement by superimposing a wavelength-offset solar spectrum on a stellar observation via the background sky spectrum.
Observations of Sun-like stars are sensitive to this effect, according to simulations lunar contamination of spectra in \citet{Roy2020}, depending on their radial velocity (and the Earth's barycentric velocity).
However, based on their results we conclude that all the observations used in this paper should have systematic errors in centroid measurements of no more than \SI{\sim2}{\meter\per\second}.

\textbf{Cosmic rays:} Excess flux in the central seven pixels of absorption features deposited by cosmic rays (or any other random source) can change the shape of the feature and thus its measured wavelength.
Rather than attempt to perform cosmic ray cleaning ourselves, we use the method described in \ref{section:pair_separation_offsets} to rejecting outliers via sigma-clipping.
While this will remove large shifts, it is likely that smaller shifts remain below the significance levels chosen.
These shifts are small by design, random, and average out over many observations, so any contribution to systematic error should be negligible.

\textbf{Stellar activity:} Changing magnetic activity in stars could potentially change the shapes of absorption features and introduce additional scatter in pair separation measurements.
Two of the solar twins in this paper, HD~45184 and HD~146233, have known activity periods shorter than the baseline of the $>100$ observations we have for them.
We visually inspected plots of the model-corrected pair separations of the 17 pairs (and their component transitions) covered in this paper, and found no evidence of any additional scatter beyond the expected statistical variation of \SI{\sim15}{\meter\per\second}.

\textbf{Effects from planets:} While achromatic Doppler shifts in a star's spectrum due to planets should not affect pair separations, transiting planets may pose a unique problem by selectively blocking red- or blue-shifted light as they transit.
This selective blocking of a star's disk could potentially change the shapes of absorption features through the Rossiter-McLaughlin effect \citep{Rossiter1924, McLaughlin1924}.
Although no solar twins in the sample are known to have transiting planets\footnote{Barring the Sun, but none of the inner planets were transiting as observed via Vesta when the observations we use were taken.}, we inspected two stars with known transiting planets (HD~39091 and HD~136352) in the larger sample of \citetalias{Berke2022b} in a similar manner as described in the previous paragraph and saw no evidence for additional scatter, at the same \SI{\sim15}{\meter\per\second} level.

\section{Results} \label{firstpaper:section:results}

\subsection{Transition pair velocity separations relative to the model: pair model offsets} \label{section:results:pair_velocity_offsets}
\begin{figure}
  \begin{center}
    \includegraphics[width=\linewidth]{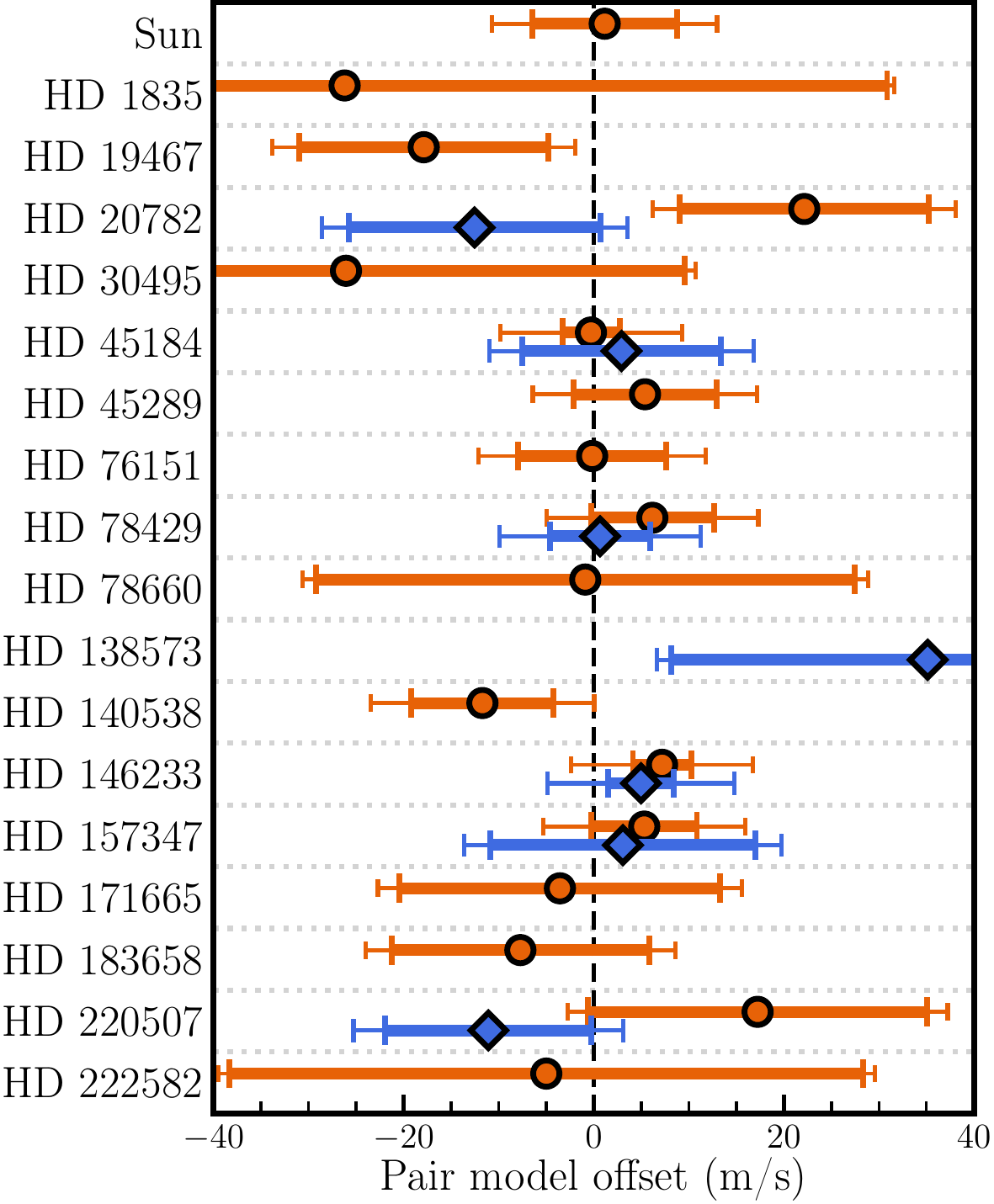}
    \caption{
    Pair model offsets for a single pair (\ion{Na}{i}\,6162.452 -- \ion{Fe}{i}\,6175.044) for all 18 solar twins separated by dotted horizontal lines.
    Results from before (orange circles) and after (blue diamonds) the HARPS fibre change are shown separately; some stars only have observations in one era.
    The thick error bar shows the statistical uncertainty, while the thin error bar shows the quadrature sum of the statistical uncertainty and the \sigsys value (for the appropriate era) for this pair.
    The weighted mean and error for the pre and post observations are $1.4\pm3.4$ and $0.61\pm\SI{5.1}{\meter\per\second}$, respectively.
    This figure is an except from the main results plot shown in \ref{figure:full_results} for the single pair.}
    \label{figure:results_excerpt}
  \end{center}
\end{figure}

\begin{landscape}
\begin{figure}
  \begin{center}
    \includegraphics[width=\linewidth]{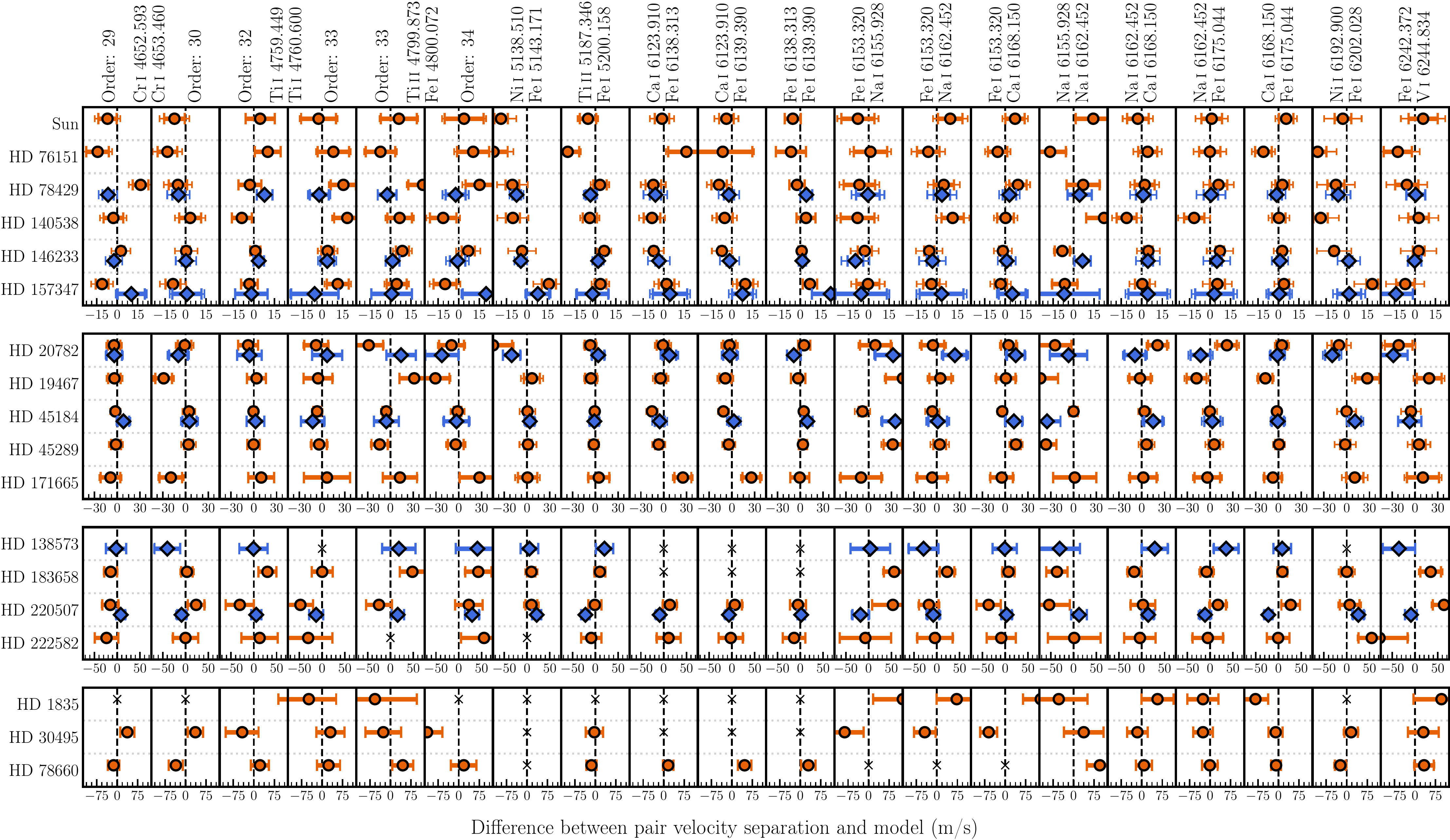}
    \caption{
    Pair model offset values for the 17 pairs and 18 solar twins considered in this paper.
    Each column shows model offsets from a single pair, with stars listed across rows.
    Results from before the HARPS fiber change are plotted as orange circles, those from after with blue diamonds.
    Some stars (generally those with a single observation) may not have any usable measurements of a particular pair's separation; these cases are marked with a black cross.
    Offsets and errors span a large dynamic range (this is primarily due to number of observations for a star), so to show as much detail as possible the stars are separated into four groups (panels) with increasing horizontal scale going down the figure.
    The thick error bar shows the purely statistical errors, while the thin error bar shows the quadrature sum of the statistical error and the \sigsys value for each pair (though as the statistical errors get larger the thin error bars may not be visible in the lower panels).
    The first three pairs have two duplicate instances each, and each instance measured and analysed separately.
    The physical order number is listed above each column for these three pairs.
    One particular point for the star HD~1835 would have required greatly expanding the scale of the fourth panel to show, harming readability; instead we note its value here: $173.2\pm\SI{83.8}{\meter\per\second}$ (2.1$\sigma$ significance) for the pair \ion{Ti}{i} 4759.449 -- \ion{Ti}{i} 4760.600 on order 32.
    }
    \label{figure:full_results}
  \end{center}
\end{figure}
\end{landscape}

The primary result of this work is the set of `pair model offsets' for the 17 transition pairs for which we have $q$-coefficients.
We use the term `pair model offset' as shorthand for the difference between the \emph{measured} separation of a pair of transitions in a star and its \emph{predicted} separation based on the models described in \ref{section:pair_separation_offsets}.
\ref{figure:results_excerpt} shows the pair model offsets for the pair \ion{Na}{i}\,6162.452 -- \ion{Fe}{i}\,6175.044 from the solar twins considered in this paper.
Each point is the weighted mean of all observations for the star listed on the left, split into observations from before (`pre') and after (`post') the HARPS optical fibre change in 2015.
\ref{figure:results_excerpt} is an excerpt from \ref{figure:full_results} which shows the results from all pairs in the sample, but we discuss what can be deduced from a single pair here before proceeding to the full results.

The first notable feature of the points in \ref{figure:results_excerpt} is their consistency.
The weighted mean values of the pre and post observations are consistent both with each other and zero within their respective errors.
The $\chi^2_\nu$ values for the two sets of observations, when considering just the statistical errors, are 1.07 and 0.97, respectively.
Visually, no additional scatter is seen among the points, implying no variation in $\alpha$ among the stars regardless of the values of the calculated $q$-coefficients.

The range of error bar sizes seen in \ref{figure:results_excerpt} is a result of several factors.
Differences in SNR between observations are one, but not likely a large factor, as the range of SNRs used is not large, between 200 (150 for the Sun) and 400; the change in statistical error due to SNR is thus at most around a factor of two.
Each point represents the weighted mean of all observations of a star in a given era, so the number of observations is a larger factor in the statistical uncertainties.
Stars with a single observation, such as HD~1835 and HD~222582, have much larger error bars than a star like HD~146233, which has 87 and 52 observations in the pre and post eras, respectively.
For the pair shown in \ref{figure:results_excerpt} the mean error (statistical and systematic) in the pre and post distributions is 19.7 and \SI{15.7}{\meter\per\second}, respectively.

\ref{figure:results_excerpt} also provides an example of the importance of the \sigsys term.
For this pair the \sigsys values are very similar between pre and post, at \SI{9.1}{\meter\per\second} and \SI{9.2}{\meter\per\second} respectively.
Without adding the \sigsys term in quadrature to the statistical errors, the pre distribution has a $\chi^2_\nu$ value of 1.07, decreasing to 0.45 with its addition; similarly, the post distribution drops from a $\chi^2_\nu$ value of 0.97 to 0.51 once the \sigsys term is incorporated.
We note that \sigsys values are derived using the larger stellar sample of 130 stars from \citetalias{Berke2022b} of which these solar twins are a subset, and that, for this particular pair, the larger sample shows greater overall scatter than the solar twins considered here.
When considering the entire sample, adding the \sigsys values in quadrature brings the $\chi^2_\nu$ of the pair model offsets for this pair down to unity; when considering just the solar twins, with their lower scatter relative to the larger sample, it reduces it to below unity.

Finally, in terms of precision, the ensemble weighted mean and uncertainty values for the `pre' and `post' distributions are $1.4\pm3.4$ and $0.61\pm\SI{5.1}{\meter\per\second}$ respectively.
The higher precision in the former is likely due to the relative number of observations in each era (306 and 111 respectively).
This level of precision is within an order of magnitude of HARPS's long-term stability of \SI{\sim1}{\meter\per\second} using cross-correlation of the entire spectrum \citep{Mayor2003}, and is an expected level of precision from this simple estimation.
We note that careful accounting for systematic errors in individual features, such as performed in this work, may be able to assist in reaching even greater precision from HARPS in other areas, such as its primary goal of planet-hunting \citep{Dumusque2018}.

\ref{figure:full_results} shows the full results for the pairs and solar twins studied in this paper.
The same broad conclusions can be drawn from it as from the excerpt in \ref{figure:results_excerpt}.
The stars shown are in good agreement with each other for all pairs, showing that the solar twins method is robust.
Some pairs show larger scatter or uncertainties on average than others, but no star consistently shows outliers across multiple pairs that might be an indication of a variation in $\alpha$.
We emphasise again that this consistency and the constraints on \varalpha it implies are independent of the exact values of the $q$-coefficients used.
Error bar size is still primarily influenced by number of observations of a star (note that the stars in the bottom panel with the largest horizontal scale all have only one or two observations).
Some star/pair combinations have no measurements that were not flagged as outliers; usually these are stars with just one or two observations.
For the full process of conversion of these results to constraints on \varalpha, see \citet{Murphy2022b}, where the ensemble precision from these 17 pairs and 18 stars is at the level of $\approx$12 parts-per-billion, almost two orders of magnitude more precise than current astronomical constraints \citep{Murphy2022}.

\begin{figure}
  \begin{center}
    \includegraphics[width=\linewidth]{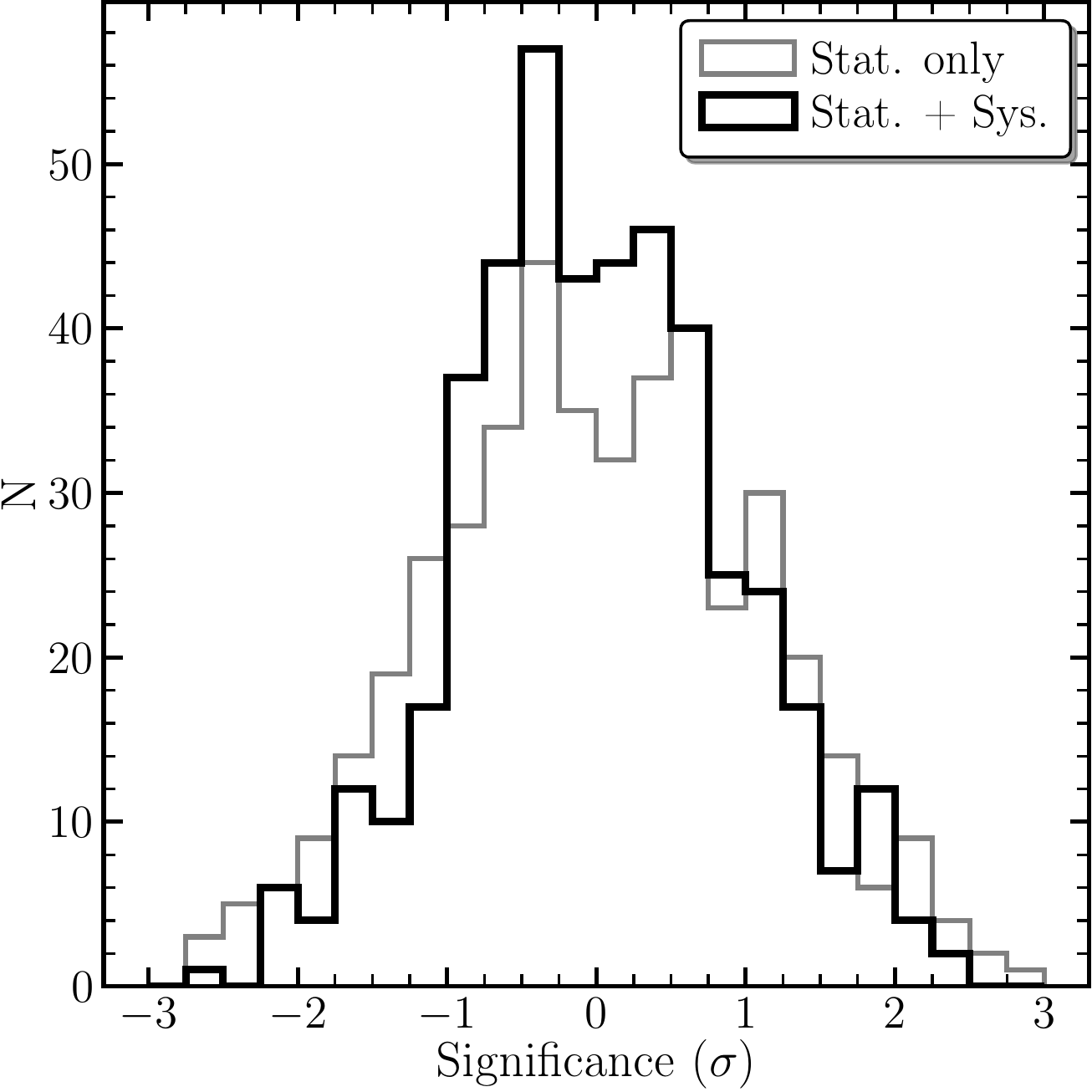}
    \caption{The distribution of the significance of pair model offsets for all stars and pairs considered in this paper, i.e., the significance values of the points in \ref{figure:full_results}.
    `Stat.' refers to the significance using just the statistical errors, while `Stat.\,+\,Sys.' the significance using the quadrature sum of statistical errors and the relevant \sigsys value for each pair.
    Here the values from before and after the HARPS optical fiber change (\ref{firstpaper:section:harps_fiber_change}) are included in each distribution.}
    \label{figure:offsets_histograms}
  \end{center}
\end{figure}

\ref{figure:offsets_histograms} shows histograms of the significance levels of all the pair measurements shown in \ref{figure:full_results}.
The lighter, thinner line shows the significance values without the appropriate \sigsys values added in quadrature, while the darker line shows the significance value with \sigsys included.
While both histograms roughly approximate Gaussian functions, the distribution without \sigsys added shows a flatter, broader profile with a small excess of scatter in the wings.
For the distribution with \sigsys added in quadrature, 74\% of values fall within 1-$\sigma$ and 97\% within 2-$\sigma$, demonstrating consistency with a Gaussian distribution.
The lack of outliers in the final results demonstrates the soundness of the solar twins method and our care in excluding spurious measurements (\ref{section:pair_separation_offsets}) which might have biased the results.
There is also no motivation for any additional clipping, with just 3\% of values having greater than 2-$\sigma$ significance in the final result.

\subsection{Origin of additional star-to-star scatter} \label{section:star_to_star_scatter}
\begin{figure}
  \begin{center}
    \includegraphics[width=\linewidth]{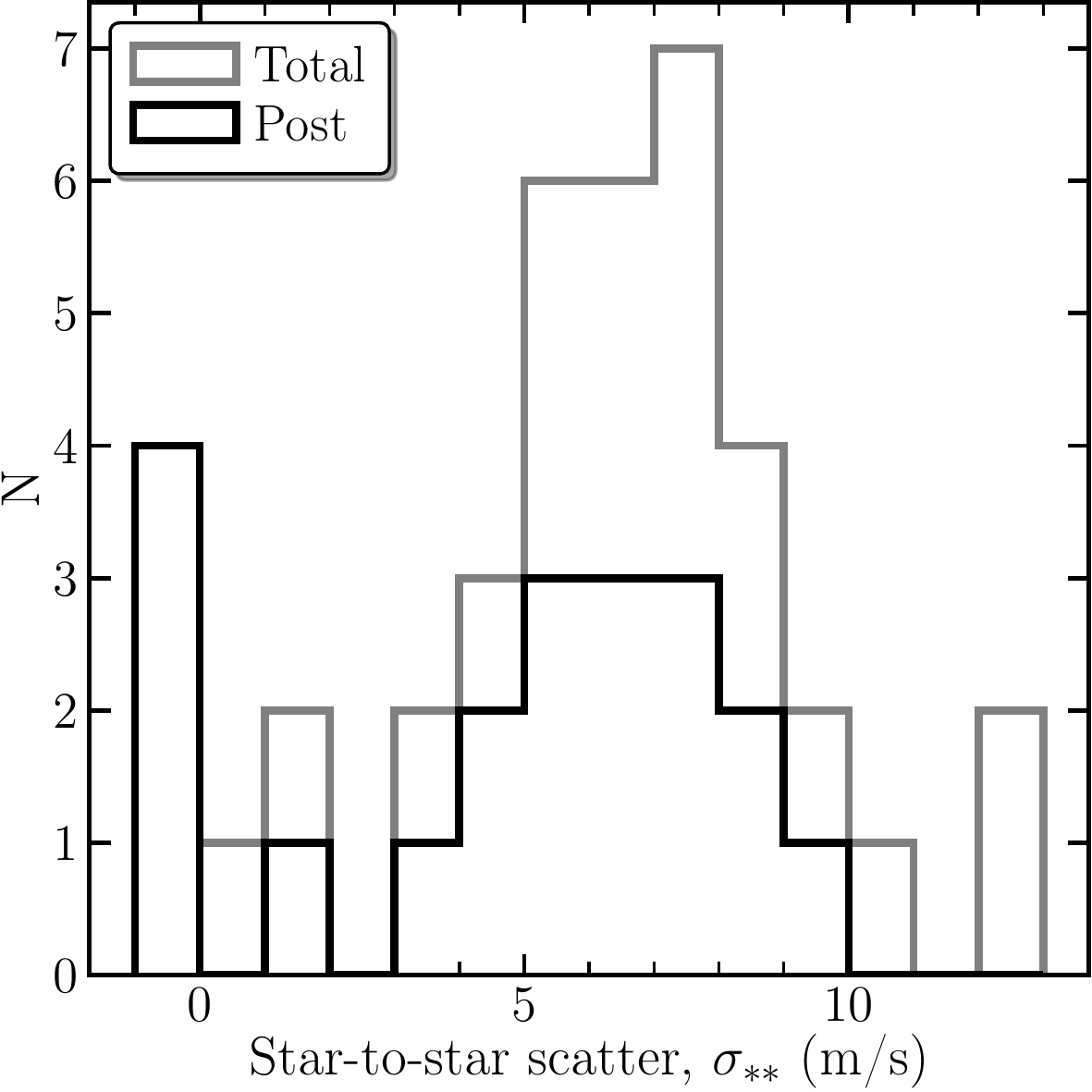}
    \caption{The distribution of \sigsys values for the 17 pairs considered in this paper.
    Three of the pairs have two instances, so there are a total of 20 \sigsys values.
    The left-most bin -- though visually to the left of zero -- counts pairs where \sigsys is \emph{equal} to zero, in a manner which gives them equal visual weight with the rest of the plot.
    A value of $\sigsys=0$ indicates no additional scatter beyond that accounted for by the statistical uncertainties was necessary to bring a pair's $\chi^2_\nu$ down to unity, i.e. its scatter could be considered purely statistical.
    The total distribution is shown as the grey histogram; values from after the HARPS fibre change (\ref{firstpaper:section:harps_fiber_change}) are shown as a black histogram to illustrate that \sigsys values of zero are only found in the post-change era.
    The medians of the pre and post distributions are 7.0 and \SI{5.6}{\meter\per\second}, respectively.
    }
    \label{figure:sigma_s2s_histogram}
  \end{center}
\end{figure}

\ref{figure:sigma_s2s_histogram} shows a histogram of \sigsys for the pairs in the sample, including separate values for the two instances in which three pairs appear.
This value, \sigsys, is an additional systematic error term which, when added in quadrature to the statistical uncertainty of a pair, brings the $\chi^2_\nu$ for that pair's model offset across all stars down to unity (\ref{section:pair_separation_offsets}).
Importantly, both observational eras (pre- and post-fibre change) show \sigsys values peaking roughly in the 5--\SI{10}{\meter\per\second} range.
That is, we find that \sigsys is non-zero for all pairs prior to the fibre change and the majority of pairs afterwards; this is discussed in greater detail in \citetalias{Berke2022b}.
Four pairs in the post-change era appear in the left-most bin where $\sigsys=0$, indicating that no extra scatter term was necessary for those pairs.

A similar plot to \ref{figure:sigma_s2s_histogram} in \citetalias{Berke2022b} (Figure 12), using a larger sample of pairs, bears out the conclusions drawn from the 17 pairs used in this paper.
However, the larger sample size of \citetalias{Berke2022b} also allowed further conclusions to be drawn, which we briefly summarise here.
In the \citetalias{Berke2022b} sample the trend of pairs needing no additional scatter term in the post-change era is seen as well, with 27 pairs (out of 284) having $\sigsys=0$.
We surmised from this result that the scatter term is likely made up of both instrumental and astrophysical components.
Errors in wavelength calibration, especially those deriving from location of measurement on the CCD, as described in \ref{firstpaper:section:harps_calibration}, could increase the scatter in pair separations.
\citet{LoCurto2015} reported decreased scatter in radial velocity measurements from cross-correlation after HARPS's 2015 upgrade, 
and the general decrease in \sigsys seen afterwards is consistent with some of the scatter in the pre-change distribution coming from instrumental variance in calibration which was removed with the fibre change.
We conclude that at least some of the extra scatter measured in the pre-change era was due to HARPS calibration inaccuracy which has since been improved.

However, for the majority of pairs (75\% here and $>90\%$ in the larger sample of \citetalias{Berke2022b}), even after the HARPS upgrade there remains a degree of star-to-star scatter beyond the statistical error bars.
Though we cannot completely exclude the possibility of further minor instrumental effects, the remaining scatter is more likely astrophysical in nature.
Two possible causes lie in differences in elemental abundances or isotope ratios between stars.
Either possibility could cause small changes in blending of absorption features (even in stars with otherwise-identical atmospheric parameters), which in turn could increase the scatter of pairs.
In \citetalias{Berke2022b} we give a more thorough description of these possible sources of scatter.
We note that this star-to-star scatter represents the dominant contributor to systematic error for this work.
Characterising \sigsys will thus be important for future improvements on the precision achieved in this paper.

\subsection{Injection and recovery of a fake signal}
We tested our ability to detect $\alpha$-variation by injecting a known signal into half of the sample stars, and seeing how well we could recover it.
For transitions with $q$-coefficients, a signal equivalent to a given \varalpha was applied to the wavelength of each transition initially measured from the spectra.
The analysis process described in \ref{firstpaper:section:analysis} was then applied to the modified data exactly as for the original data.
This allowed us to test whether our subsequent processing of these measurements successfully recovers a \varalpha signal.
We found that the introduced signal was well-recovered, and is transformed into a \varalpha value as described in \citet{Murphy2022b}, where the signal is transferred from a velocity-space measurement to a measurement of \varalpha.
When a large signal was introduced, equivalent to $\varalpha=150$ parts-per-billion (ppb), this created outliers which were rejected in our transition measurement process (\ref{firstpaper:section:automated_fitting}).
In this case, only $\sim70\%$ of the signal was recovered.
Using smaller, but still large, injected signals allowed a larger proportion to be recovered, for example, $\sim86\%$ of the signal was recovered for a shift equivalent to $\varalpha=100$ ppb.
We point out that a signal of 100 ppb is still an order of magnitude smaller than the precision of the current best astronomical constraints on \varalpha \citep{Murphy2017}.

These tests helped us refine the transition offset measurement process, and allowed us to determine that a $3\sigma$ clipping limit for transitions was the best compromise (cf. $2.5\sigma$ or $3.5\sigma$).
For small signals injected in this way, some fraction of the signal will always be unrecoverable due to the model adjusting slightly during the process of fitting the transition offsets and pair separations.
If instead the measurements of this paper are used as a reference against which a more distant stellar sample is compared, our signal injection test demonstrates that even such small signals will be recovered faithfully.

\section{Conclusions} \label{firstpaper:section:conclusions}
In this paper we provide details of the solar twins method introduced in \citet{Murphy2022b}, and have applied the method to the Sun and 17 of the most Sun-like stars (solar twins, according to \ref{equation:star_definitions}) in the extensive HARPS archive.
A total of 423 HARPS spectra with $\text{SNR}\geq200$ (150 for solar spectra) were used.
Transitions were carefully selected to avoid telluric features (down to 0.1\% of the continuum) and categorised based on blending with other stellar features, with only the least-blended 164 features used.
Based on our analysis of systematic error with blending we conclude that our selection of transitions is conservative, and contributes no significant error.
Our analysis involved fully automated fitting of these lines, and correction of several instrumental artefacts.
\ref{firstpaper:section:summary_of_systematic_errors} contains a summary of the systematic effects investigated; a detailed description of each is provided in companion paper \citet{Berke2022b} (\citetalias{Berke2022b}).
We modelled the separations between 17 pairs of transitions with $q$-coefficients (sensitivities to $\alpha$-variation) as functions of the stellar parameters \Teff, [Fe/H], and $\log{g}$; the resulting models are robust enough to be able to serve as the reference against which solar twins are compared.
These pair separations, compared to the reference models, then provide constraints on \varalpha without any reference to laboratory wavelength measurements.

The most important result of this work is that solar twins can serve as precise and accurate probes for searches of $\alpha$-variation.
The solar twins method provides a completely differential measurement which can sensitively detect variation in pair separation between stars.
This work establishes a local reference sample ($<\SI{50}{pc}$) to which any similar future work can be compared.
No evidence is seen, for any of the 17 transition pairs in any of the 18 stars, of a change in pair separation which might indicate a variation in $\alpha$, demonstrating the robustness of the method.
The 17 pairs have ensemble precisions ranging from 2.0 to \SI{6.2}{\meter\per\second} with a median of \SI{3.5}{\meter\per\second}, from  310 and 113 observations taken before and after the HARPS fibre change, respectively.
For comparison, the uncertainty on individual pair separation measurements is typically $\sim\SI{30}{\meter\per\second}$.
To provide a simple example, for a pair with a difference of $q$-coefficients of $\sim$\SI{1500}{\per\centi\meter}, an ensemble precision of \SI{3}{\meter\per\second} translates to a precision in \varalpha of $\sim67$ parts-per-billion (ppb), or $\sim16$ ppb when averaged over the 17 pairs in the sample.
\citet{Murphy2022b} presents the full details of the conversion process and the resulting formal constraints on \varalpha, which they find to be about 12 ppb.
These results are nearly two orders of magnitude more precise than the current most precise astrophysical measurements \citep[see e.g.,][]{Murphy2017}.

An important additional finding is the existence, for most pairs, of star-to-star scatter in pair separations.
This value, \sigsys, is the excess scatter beyond that expected from the statistical errors and ranges from 0 to \SI{12}{\meter\per\second}.
The median values, for observations before and after the HARPS fibre change in 2015, are 7.0 and \SI{5.6}{\meter\per\second}, respectively.
This is currently the largest systematic error identified in our analysis, but we note that averaging over many stars reduces its importance for the ensemble results.
Bolstered by the larger stellar sample of 130 stars from \citetalias{Berke2022b} (which extends the solar twins method to solar analogues\footnote{With $\Teff\pm\SI{300}{\kelvin}$, $\text{[Fe/H]}\pm0.3$, and $\log{g}\pm0.4$ around solar values, parameter ranges two or three times larger than for solar twins.}), we conclude that \sigsys is likely composed of both an instrumental and an astrophysical component.
The fact that 25\% of pairs in the post-fibre change era require no extra scatter (unlike in the pre-change era) suggests that at least part of the scatter is due to instrumental effects.
This is consistent with the wavelength calibration of individual lines in HARPS changing after the fibre change, with a small increase reported in overall radial velocity stability \citep{LoCurto2015}.
However, the fact that the overall shape of the distribution of \sigsys values remains the same after the change for 75\% of pairs ($>90\%$ in \citetalias{Berke2022b}), with a peak clearly separated from zero, suggests that there is also an astrophysical component to it.
\ref{section:star_to_star_scatter} discusses possible sources for this scatter, though the relatively small size of our stellar sample in this paper prevents us from making firm conclusions; however, based on the results from the larger sample of 130 Sun-like stars, we concluded in \citetalias{Berke2022b} that it can likely be explained by differences in stellar rotation speeds and inclinations, and/or abundance and isotope ratios between stars.

The ability to use main-sequence stars as probes of \varalpha opens the window to much more precise searches for $\alpha$-variation on the Galactic scale.
With data from the \textit{Gaia} mission we are poised to discover solar twins at much greater distances (up to \SI{4}{kpc}) than the ones in our sample, with initial progress on this front being reported in \citet{Lehmann2022}.
By extending the usable range of all three atmospheric parameters to include solar analogues, \citetalias{Berke2022b} allows a larger number of stars to be used, which helps lower the stellar parameter precision requirements in distant target-finding campaigns.
Finding distant solar twins deeper in the Galactic dark matter halo would provide a prime test bed for proposed extensions to the Standard Model such as the possibility of variation in $\alpha$ as a function of dark matter density \citep[e.g.,][]{Olive2002, Davoudiasl2019}.
Further progress will be reported in Lehmann et al. (in prep.) and \textcolor{black}{Liu et al. (in prep.)}.
More distant solar twins will inevitably provide lower SNR spectra, but the analysis here demonstrates that, while the statistical errors will be larger, the method will continue to provide precise and accurate results.

Finally, the twins technique detailed here could potentially be extended to other spectral classes of stars.
While solar twins demonstrably make good probes of 
\varalpha, other intrinsically more luminous spectral types of stars (e.g. giants) may prove to be usable as well.
Additional types of stars might offer the ability to probe $\alpha$-variation on even larger distance scales, or act as an additional consistency check for solar twins.

\subsection*{Acknowledgements}
We thank Dainis Dravins for discussions about potential astrophysical systematic errors.
We would also like to thank the anonymous referee whose comments led to significant improvements in the paper.

\textsl{Funding:} DAB, MTM and FL acknowledge the support of the Australian Research Council through \textsl{Future Fellowship} grant FT180100194.

\textit{Facilities.} This research has made use of data or services obtained from, or tools provided by: the Extrasolar Planets Encyclopaedia (Jean Schneider, CNRS/LUTH - Paris Observatory), the SIMBAD database, operated at CDS, Strasbourg, France, NASA's Astrophysics Data System, the Geneva--Copenhagen survey \citep{Nordstrom2004}, the Belgian Repository of fundamental Atomic data and Stellar Spectra \citep{Lobel2008}, atomic line lists compiled by R. Kurucz, the NIST Atomic Spectra Database \citep{Kramida1999, Kramida2013}, the Transmissions Atmosph\'eriques Personnalis\'ees Pour l'AStronomie project \citep{Bertaux2014}, and the ESO Science Archive Facility.

\textit{Software.} This research has made use of: Python \citep{VanRossum1995}, NumPy \citep{Harris2020}, SciPy \citep{Virtanen2020}, Astropy \citep{Astropy2013, Astropy2018}, Astroquery \citep{Ginsburg2019}, CMasher \citep{vanderVelden2020}, IPython \citep{Perez2007}, \texttt{matplotlib} \citep{Hunter2007}, \texttt{unyt} \citep{Goldbaum2018}, \texttt{hickle} \citep{Price2018}, \texttt{tqdm} \citep{DaCostaLuis2021}, TOPCAT \citep{Taylor2005}, \texttt{ds9} \citep{Joye2003}, and \texttt{pandas} \citep{McKinney2010, McKinney2011}.

\subsection*{Data Availability}
Based on observations obtained from the ESO Science Archive Facility and collected at the European Southern Observatory under ESO programme(s) 188.C-0265(C), 188.C-0265(E), 188.C-0265(G), 188.C-0265(J), 188.C-0265(K), 188.C-0265(O), 188.C-0265(Q), 075.C-0332(A), 072.C-0488(E), 077.C-0364(E), 183.D-0729(A), 183.C-0972(A), 192.C-0852(A), 196.C-1006(A), 198.C-0836(A), 60.A-9036(A), 188.C-0265(A), 188.C-0265(P), 188.C-0265(R), 188.C-0265(B), 188.C-0265(M), 091.C-0936(A), and 088.C-0323(A).
The raw data for this work is available from the ESO Science Archive facility, at \url{http://archive.eso.org/cms/eso-data/eso-data-direct-retrieval.html}.




\appendix
\section{Transitions used in this paper}
\ref{table:transitions} contains a list of the transitions with $q$-coefficients used in this paper, along with information on the pairs formed from the transitions.

\bsp 

\begin{landscape}
\begin{table}
    \caption{
    Transitions with $q$-coefficients used in this paper.
    The first three columns show the vacuum wavelength and wavenumber \citep{Kramida1999} of the transition and which species it comes from.
    The next two sets of three columns show the energy, full orbital configuration term, and angular momentum J of the lower and upper states of the transition, respectively.
    The next two columns show the $q$-coefficient and its uncertainty (not a formal statistical uncertainty) for the transition, as calculated in \citet{Dzuba2022}.
    The penultimate column gives the number of the pairs which each transition is part of; there are 17 pairs in total, and some transitions appear in multiple pairs.
    The final column, labeled ``Bl.'', is the ``blendedness'' rating of each transition.
    }
    \centering
    \begin{tabular}{|p{1.7cm}p{1.5cm}l|p{1.4cm}lr|p{1.4cm}lr|>{\raggedleft}p{1cm}>{\raggedleft}p{1.2cm}|l|l|}
    \hline
       Wavelength (\AA, vacuum) & Wavenumber (\si{\per\centi\meter}) & Ion & Energy (\si{\per\centi\meter}) & Orbital configuration (lower) & J & Energy (\si{\per\centi\meter}) & Orbital configuration (upper) & J & $q$ (\si{\per\centi\meter}) & d$q$ (\si{\per\centi\meter}) & Pairs & Bl. \\
    \hline
    4652.593 & 21493.391 & Cr I  &  7927.441 & 3d4 4s2 a 5D         & 2     & 29420.864    & 3d4 (5D) 4s 4p (3P*) y 5P*     & 1    &     490 &   60 & 1 & 1 \\
    4653.460 & 21489.386 & Cr I  &  8095.184 & 3d4 4s2 a 5D         & 3     & 29584.571    & 3d4 (5D) 4s 4p (3P*) y 5P*     & 2    &     490 &   60 & 1 & 1 \\
    4759.449 & 21010.835 & Ti I  & 18141.265 & 3d3 (2H) 4s a 3H     & 5     & 39152.103    & 3d3 (2H) 4p x 3H*              & 5    &     440 &   60 & 2 & 0 \\
    4760.600 & 21005.756 & Ti I  & 18192.570 & 3d3 (2H) 4s a 3H     & 6     & 39198.323    & 3d3 (2H) 4p x 3H*              & 6    &     460 &   60 & 2 & 0 \\
    4799.873 & 20833.885 & Ti II &  8710.567 & 3d2 (1D) 4s a 2D     & 3/2   & 29544.454    & 3d2 (3F) 4p z 4G*              & 5/2  &      90 &   90 & 3 & 2 \\
    4800.072 & 20833.021 & Fe I  & 12968.554 & 3d7 (4F) 4s a 3F     & 2     & 33801.572    & 3d7 (4F) 4p y 5D*              & 2    &  $-830$ &  400 & 3 & 2 \\
    5138.510 & 19460.894 & Ni I  & 13521.347 & 3d8 (1D) 4s2 1D      & 2     & 32982.260    & 3d9 (2D) 4p 1P*                & 1    &    4950 &  500 & 4 & 2 \\
    5143.171 & 19443.258 & Fe I  & 19552.478 & 3d6 4s2 a 3P2        & 1     & 38995.736    & 3d7 (4F) 4p y 3D*              & 1    &    2840 &  300 & 4 & 1 \\
    5187.346 & 19277.681 & Ti II & 15265.700 & 3d2 (1G) 4s b 2G     & 7/2   & 34543.380    & 3d2 (3F) 4p z 2G*              & 7/2  &     340 &  150 & 5 & 1 \\
    5200.158 & 19230.185 & Fe I  & 17927.382 & 3d7 (4P) 4s a 5P     & 1     & 37157.566    & 3d6 (5D) 4s 4p (1P*) y 5P*     & 2    & $-1500$ &  900 & 5 & 1 \\
    6123.910 & 16329.437 & Ca I  & 15210.063 & 3p6 4s 4p 3P*        & 1     & 31539.495    & 3p6 4s 5s 3S                   & 1    &   $-69$ &    7 & 6,7 & 1 \\
    6138.313 & 16291.121 & Fe I  & 19788.251 & 3d6 4s2 a 3H         & 4     & 36079.372    & 3d7 (4F) 4p z 3G*              & 3    &    2710 &  500 & 6,8 & 2 \\
    6139.390 & 16288.263 & Fe I  & 20874.482 & 3d6 4s2 b 3F2        & 3     & 37162.746    & 3d7 (4F) 4p y 3F*              & 3    &    3010 &  500 & 7,8 & 2 \\
    6153.320 & 16251.389 & Fe I  & 17550.181 & 3d7 (4P) 4s a 5P     & 3     & 33801.572    & 3d7 (4F) 4p y 5D*              & 2    &     340 &  400 & 9,10,11 & 0 \\
    6155.928 & 16244.504 & Na I  & 16956.170 & 2p6 3p 2P*           & 1/2   & 33200.673    & 2p6 5s 2S                      & 1/2  &       7 &    1 & 9,12 & 2 \\
    6162.452 & 16227.307 & Na I  & 16973.366 & 2p6 3p 2P*           & 3/2   & 33200.673    & 2p6 5s 2S                      & 1/2  &   $-11$ &    1 & 10,12,13,14 & 1 \\
    6168.150 & 16212.316 & Ca I  & 20335.360 & 3p6 3d 4s 3D         & 1     & 36547.688    & 3p6 4s 5p 3P*                  & 0    &  $-475$ &   24 & 11,13,15 & 1 \\
    6175.044 & 16194.217 & Fe I  & 17927.382 & 3d7 (4P) 4s a 5P     & 1     & 34121.603    & 3d7 (4F) 4p y 5D*              & 0    &     430 &  500 & 14,15 & 1 \\
    6192.900 & 16147.524 & Ni I  & 13521.347 & 3d8 (1D) 4s2 1D2     & 2     & 29668.893    & 3d9 (2D) 4p 3D3                & 3    &    5500 &  550 & 16 & 2 \\
    6202.028 & 16123.758 & Fe I  & 21038.987 & 3d6 4s2 b 3F2        & 2     & 37162.746    & 3d7 (4F) 4p y 3F*              & 3    &    2870 &  500 & 16 & 1 \\
    6242.372 & 16019.552 & Fe I  & 17927.382 & 3d7 (4P) 4s a 5P     & 1     & 33946.933    & 3d6 (5D) 4s 4p (3P*) z 3P*     & 2    & $-1300$ &  900 & 17 & 1 \\
    6244.834 & 16013.236 & V I   &  2424.809 & 3d4 (5D) 4s a 6D     & 9/2   & 18438.044    & 3d3 (4F) 4s 4p (3P*) z 6D*     & 9/2  &  $-600$ &   60 & 17 & 1 \\
    \hline
    \end{tabular}
\label{table:transitions}
\end{table}

\label{lastpage}
\end{landscape}

\end{CJK*}

\end{document}